\RequirePackage{lineno}
\documentclass{JINST}

\def\Om{\Omega}

\def\lsim{\mathrel{\rlap{\lower4pt\hbox{\hskip1pt$\sim$}}
    \raise1pt\hbox{$<$}}}
\def\gsim{\mathrel{\rlap{\lower4pt\hbox{\hskip1pt$\sim$}}
    \raise1pt\hbox{$>$}}}

\newcommand{\beq}{\begin{eqnarray}}
\newcommand{\eeq}{\end{eqnarray}}

\def\um{\mbox{m}}

\title{Testing of Cryogenic Photomultiplier Tubes for the MicroBooNE Experiment}

\author{
T.~Briese$^d$, 
L.~Bugel$^b$, 
J.~M.~Conrad$^b$, 
M.~Fournier$^d$, 
C.~Ignarra$^b$, 
B.~J.~P.~Jones$^b$, 
T.~Katori$^b$, 
R.~Navarrete-Perez$^c$\footnote{
Present address: Astronomy and Astrophysics Department, University of California, 
Santa Cruz, 95064}, 
P.~Nienaber$^d$, 
T.~McDonald$^d$, 
B.~Musolf$^e$, 
A.~Prakash$^b$\footnote{
Present address: Physics Department, The Ohio State University, Columbus, 43210}, 
E.~Shockley$^d$, 
T.~Smidt$^b$, 
K.~Swanson$^a$, 
M.~Toups$^b$\\
\llap{$^a$}Physics Department, California Institute of Technology, Pasadena, CA 91125\\
\llap{$^b$}Physics Department, Massachusetts Institute of Technology, Cambridge, MA 02139\\
\llap{$^c$}Instituto de Ciencias Nucleares, 
Universidad Nacional Aut\'onoma de M\'exico, D.F. 04510, M\'exico\\
\llap{$^d$}Physics Department, Saint Mary's University of Minnesota, Winona, MN 55987 \\
\llap{$^e$}Warren Township High School, Gurnee, IL 60031 \\
E-mail: \email{KATORI@FNAL.GOV}}

\abstract{The MicroBooNE detector, to be located on axis in 
the Booster Neutrino Beamline (BNB) at the Fermi National Accelerator Laboratory (Fermilab), 
consists of two main components: a large liquid argon time projection chamber (LArTPC), 
and a light collection system. 
Thirty 8-inch diameter Hamamatsu R5912-02mod cryogenic photomultiplier tubes (PMTs) 
will detect the scintillation light generated in the liquid argon (LAr). 
This article first describes the MicroBooNE PMT performance test procedures, including how the light collection
system functions in the detector, and the design of the PMT base.  The design of the cryogenic test stand is 
then discussed, and finally the results of the cryogenic tests are reported.}


\begin{document}

\section{Introduction}

\subsection{The MicroBooNE experiment at Fermilab}

The MicroBooNE detector is a LArTPC positioned in the BNB at the Fermilab, 
scheduled to begin data collection in  2014~\cite{uB}. 
It is a milestone experiment on 
the path to future large LArTPC detectors for  long baseline neutrino oscillation experiments, 
such as LBNE~\cite{LBNE}, LBNO~\cite{LBNO}, and Okinoshima~\cite{Okinoshima}; 
it is expected that MicroBooNE will provide important design and construction information for these experiments.
The 170 tons of liquid argon of the MicroBooNE cryostat contains $2.5\times2.4\times10.4~\um^3$ 
TPC drift volume. 
Thirty cryogenic 8-inch PMTs 
are located behind the wire planes to observe the argon scintillation light 
following the basic design of the ICARUS T600 detector~\cite{ICARUS_detec}. 

\subsection{The light collection system}

The scintillation light from a minimum ionizing particle (MIP) is 
roughly 40,000 photons/MeV~\cite{doke_absolute}. 
The presence of electric field (500~V/cm for MicroBooNE) 
reduces this to $\sim$60\%~\cite{kubota_Efield}. 
Among them, 30\% of light corresponds to the prompt light~\cite{hitachi_IsIt} 
and useful to trigger the TPC. 
Thus, a 1~MeV MIP in the MicroBooNE TPC volume 
emits $\sim$7200 prompt photons. 

The light collection system of MicroBooNE consists of 32 8-inch PMTs. 
Large photocathode PMTs provide a cost-effective solution to maximize photocathode coverage; 
this design yields a 0.9\% photocathode coverage. 
By this choice, conservatively, PMTs with $\sim$1\% total efficiency 
(including PMT quantum efficiency, 
wavelength shifter efficiency, 
and geometric loss), 
can trigger, say, a 10~MeV MIP.

The very important feature of the light collection system is its ability to 
measure the event time at the $\sim$ns level.
Since prompt light from the liquid argon is emitted on a 3-6~ns time scale, 
detection of the scintillation light allows the LArTPC to be triggered, 
where drift of electrons take $\sim$ms.
Although MicroBooNE is an accelerator-based neutrino beam experiment and 
trigger information can be obtained from the accelerator clock via the Fermilab ACNET (accelerator network), 
there are a number of reasons that the light collection system is useful.
First, any non-beam related physics, 
such as supernova neutrinos, 
requires a trigger from the light collection system. 
Second, scintillation light information can  distinguish physics events from cosmic ray backgrounds via timing.
MicroBooNE will run on the surface, 
and $\sim$kHz rates of cosmic ray events in the drift volume are anticipated.
The probability that cosmic rays will overlap with the BNB beam pulse (1.6~$\mu$s) is low, 
but any cosmic rays before or after the beam pulse still will leave tracks in the TPC 
if those rays pass during the drift time ($\sim$ms), causing a confusion in the event matching. 
Third, scintillation light itself is a useful measure of the total energy deposited by an event.
Finally, 
the pulse shape information from the scintillation light could be used for particle identification.

\subsection{The PMT unit}

\begin{figure}
\centering
\includegraphics[width=0.60\textwidth]{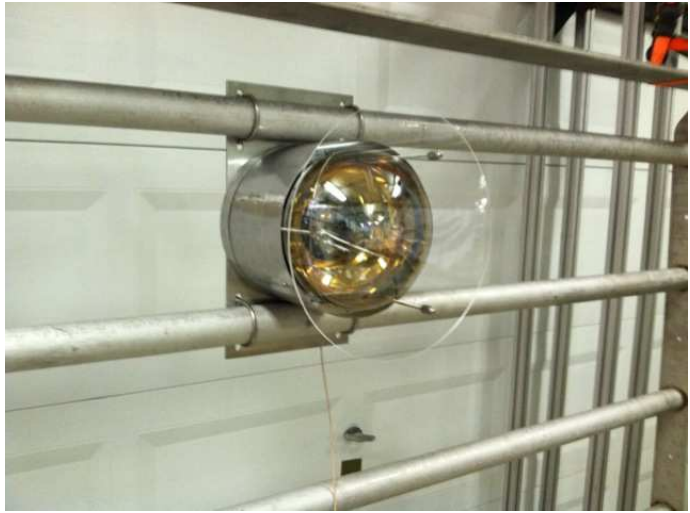}
\caption{\label{fig:PMTunit}
A picture of the mechanical model of the MicroBooNE PMT unit in the test facility. 
A PMT is fixed in the mount, which is surrounded by 
the magnetic shield up to the equator of the PMT, 
and the wave length shifter deposited acrylic plate is equipped in front of the PMT.
}
\end{figure}

Each PMT unit is defined by a PMT
in the PMT mount (Teflon mount fixed by spring loaded Teflon coated wires) 
fitted with a wavelength shifter deposited acrylic plate 
and a magnetic shield. 
Figure~\ref{fig:PMTunit} shows the prototype of the MicroBooNE PMT unit. 

The light collection system uses 8-inch diameter, bi-alkali photocathode  
Hamamatsu R5912-02mod cryogenic PMTs~\cite{Hamamatsu}. 
The R5912-02mod is a 14-stage high gain PMT
($10^9$ gain with the average optimal voltage $\sim$1700~V);
the intrinsically high gain of the tube helps to compensate for 
the diminution of the gain in the cryogenic environment.
This also permits operation at a slightly lower voltage, which avoids high voltage (HV) 
breakdown on the base, the cable connection, or the feed-through. 
The bi-alkali photocathode stops working below 150~K~\cite{Hans1}. 
This is cured by adding a thin platinum layer between the photocathode and 
the borosilicate glass envelope 
to preserve the conductance of the photocathode at low temperature. 
The large photo-cathode and high gain makes R5912-02mod a suitable choice for us. 

Since the argon scintillation light is in the vacuum UV region (128~nm) 
and thus cannot penetrate glass, 
tetra-phenylbutadiene (TPB) is used as a wavelength shifter.  A TPB-polystyrene mixture is deposited 
on an acrylic plate positioned in front of each PMT~\cite{benchmark}. 
The degradation of TPB, 
particularly by the exposure to UV light, 
has been carefully studied~\cite{Christie,Ben}. 

The performance of such a large diameter PMT is sensitive to the Earth's magnetic field and 
a cryogenic magnetic shield has been designed and tested.
It is made of a material featured with 
nonzero permeability at  cryogenic temperatures~\cite{AK4}. 
Our tests show that these specially made shields are effective at liquid nitrogen temperatures,
and shielding performance, especially with respect to the orientation of 
the PMT in the earth magnetic shield, 
is consistent with other published results~\cite{DCshield}. 

\subsection{The PMT base design}

\begin{figure}
\centering
\includegraphics[width=1.00\textwidth]{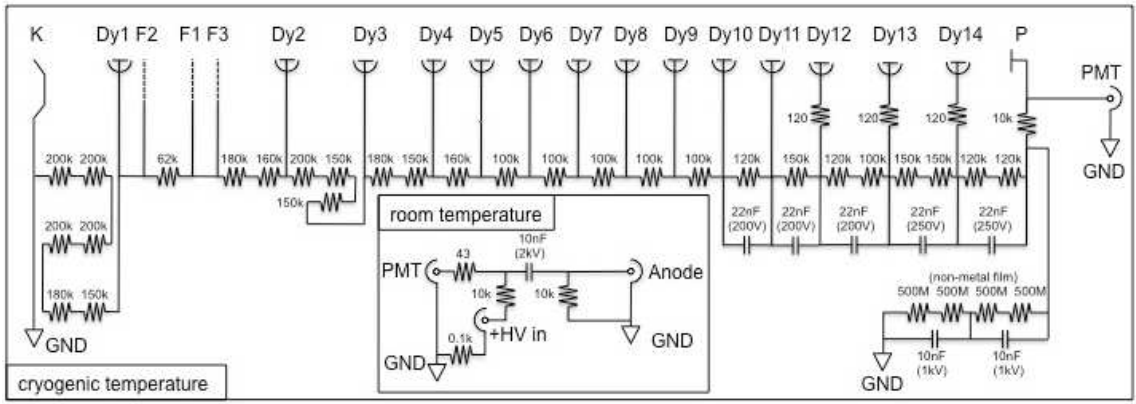}
\caption{\label{fig:schematic}
The schematic of the MicroBooNE PMT. 
PMTs and bases are both located inside the cryostat at cryogenic LAr temperature. 
Since the PMT is operated with positive HV, 
a decoupling capacitor is required to split the signal from the HV 
outside of the cryostat (that is, at room temperature). 
}
\end{figure}

The base is designed for  positive HV operation, {\it i.e.}, 
the photocathode is grounded, and the last dynode in the chain is held at high voltage. 
This scheme has a number of advantages for present  purposes. 
First, only  one cable is needed for each channel, as each cable carries both the DC HV and the signal. 
This reduces the number of penetrations of the cryostat, 
which often require special structures. 
Second, grounding the photocathode reduces noise, which helps with the detection of low energy events.
Third, a grounded photocathode will not interfere with the TPC. 
Since the PMTs are located right behind the TPC anode wires, 
a nonzero potential at the PMT photocathode would alter the TPC electric field.
For MicroBooNE, the last collection plane is grounded. 
A grounded photocathode will therefore not affect the field in the TPC field cage. 
Finally, grounding the photocathode makes for a good electric field inside the PMT.
When the photocathode is not grounded, the tube requires a grounded shield, 
which is cumbersome for the large diameter PMT MicroBooNE uses.
 
Figure~\ref{fig:schematic} shows the schematic of the MicroBooNE PMT base. 
The dynode chain follows the recommended  resistance ratios from the manufacturer~\cite{Hamamatsu}. 
Several  features were added to this basic design. 
A charge "reservoir'' at the last stage is provided by 
the large capacitors through the ground, 
and this reservoir supplies electrons quickly to the later dynodes. 
Since PMT is operated with positive high voltage, 
the signal needs to be split from the HV; this happens outside the cryostat, 
by means of a room temperature splitter circuit.

\begin{figure}
\centering
\includegraphics[width=0.80\textwidth]{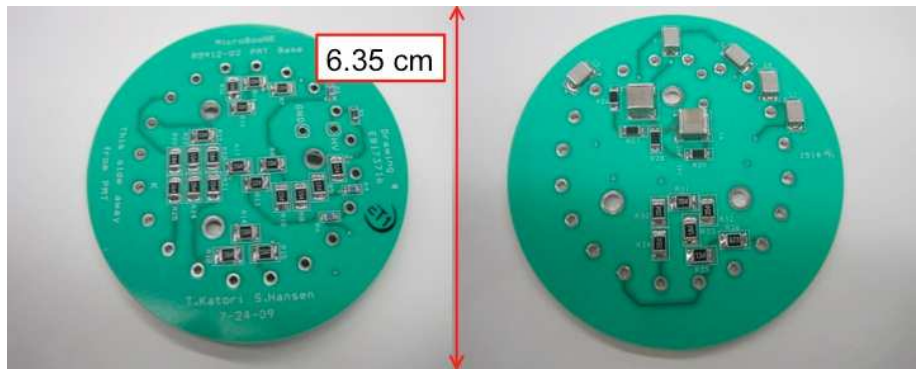}
\caption{\label{fig:base}
Pictures of the MicroBooNE PMT base. 
}
\end{figure}

Figure~\ref{fig:base} shows a picture of the MicroBooNE PMT base. 
Special attention is paid to the choice of passive components. 
All passive components are surface mounted, 
and only metal film resistors and C0G/NP0 type capacitors, 
known to have the smallest temperature dependencies, are used.
Materials for the PC board and the cable are also chosen carefully to minimize 
water and oxygen contamination, 
and to perform at cryogenic temperatures.
The PC board is made from Roger4000 series~\cite{Roger4000}, 
and the cable has an FEP outer jacket and PTFE inner insulator. 

All bases were tested after the production and were then soldered to a PMT. 
No thermal contraction caused base failures have been observed.  
A similar base was designed for a 2-inch head-on 
Hamamatsu R7725-mod cryogenic PMT and used in R\&D studies at the
Massachusetts Institute of Technology (MIT)~\cite{lightguide}. 
The MIT group observed several base failures due to 
the failure of the 10~k$\Om$ resistor at 
the end of the dynode chain. 
Damage to this component appears to have been caused by the impulse current generated 
by the splitter shorting in the gaseous phase argon.

\section{PMT test stand}

\subsection{PMT test for MicroBooNE}

Thirty-three (30 plus 3 spare)  8-inch MicroBooNE PMTs were tested 
both in air (room temperature) and in liquid nitrogen (LN2, cryogenic temperature). 
Here the test in the LN2 (77~K) represent the MicroBooNE environment, where all apparatus 
is immersed in the liquid argon (87~K). 
The main purpose of this test was to identify possibly defective PMT assemblies in a cryogenic environment 
before installing them in the cryostat.
Two main parameters were measured: the gain and the dark current. 
We found all of the PMTs performed properly at cryogenic temperature. 
A sample of the PMTs were selected for specific additional tests.

\subsection{PMT test stand design}

\begin{figure}
\centering
\includegraphics[width=0.80\textwidth]{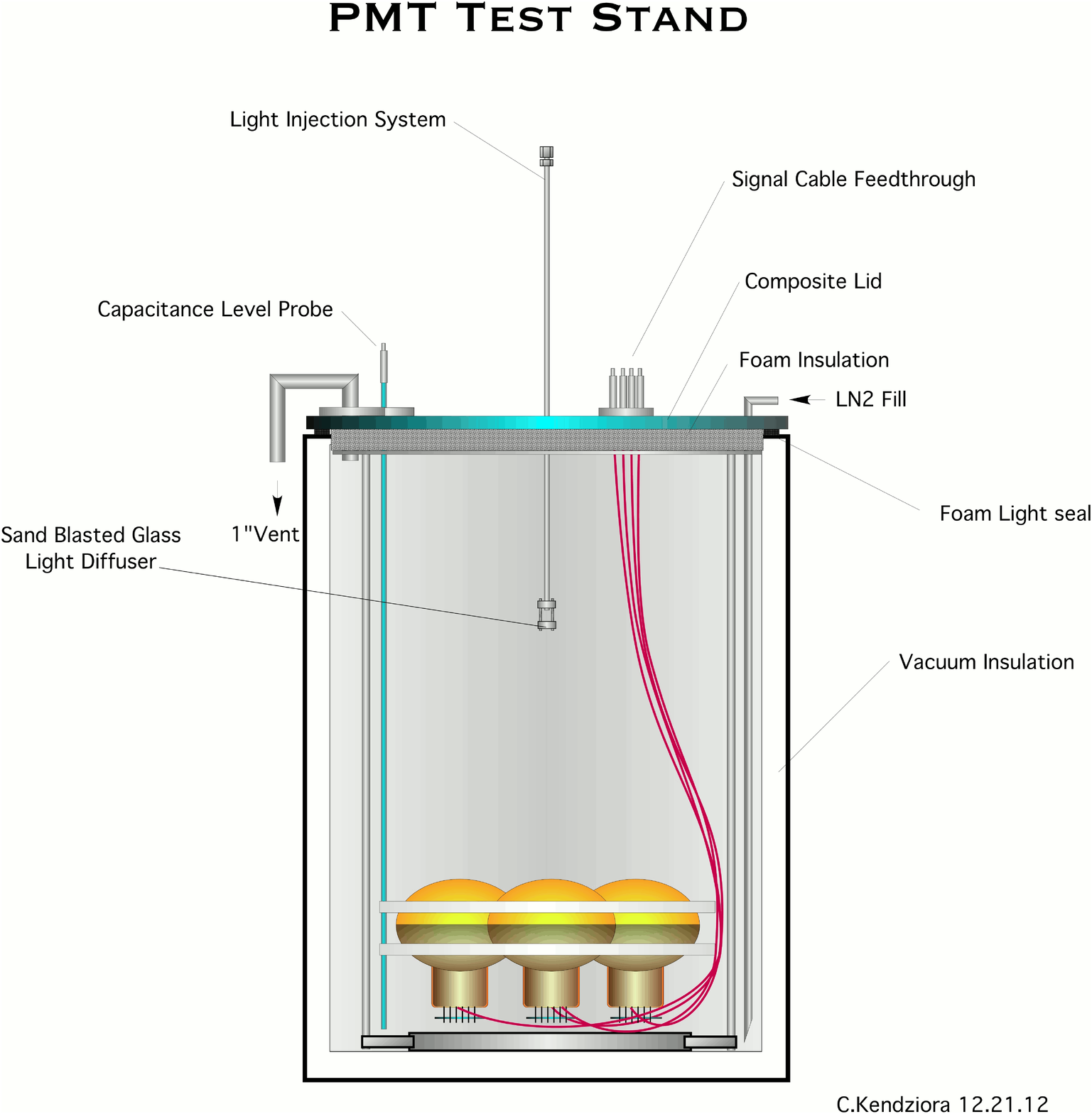}
\caption{\label{fig:teststand}
The PMT test stand. 
Four PMTs are held at the bottom of the stand; they are simultaneously tested by exposure to
 light from the optical fiber assembly. 
}
\end{figure}

Figure~\ref{fig:teststand} shows a drawing of the PMT test stand.
The stand is based on a 346~L open Dewar~\cite{dewar}, 
which can be filled either with LN2 or with room-temperature air.
The Styrofoam and glass fiber Dewar lid is modified to include five penetrations: a LN2 filling line, 
a vent for gaseous nitrogen from boil-off, a LN2 level-monitor probe, 
a cable feed-through, and a light injection system to carry measured amounts of 
light into the otherwise light-tight interior. 
The cable feed-through is a commercial four-connector SHV vacuum feed-through~\cite{feedthrough}, 
with a custom modification to prevent high-voltage breakdown (allowing the connector to be used in gaseous argon). 
An optical fiber runs inside a metal fixture passing through the center of the lid.  
The exterior end of the fiber is exposed to light from a pulsed blue LED and 
the end of the fiber inside the Dewar is optically coupled to a small sand-blasted 
glass piece which diffuses the light so as to illuminate the PMTs under test.
Four PMTs are held within a Delrin fixture, which is  
attached to the underside of the lid by a stainless steel mount.  
The PMTs sit on the bottom Delrin fixture during room temperature running, 
and tubes fit into the openings of the top Delrin fixture by the buoyancy during cryogenic running. 
This configuration avoids putting pressure on the PMTs by thermal contraction of the fixture.

\subsection{Test procedure}

In cryogenic running, the PMTs are immersed in  liquid nitrogen, 
and kept in a dark environment for at least three days
before any tests are performed, with the exception of  the cooling test (Sec.~\ref{subsubsec:cooling}). 
One calibration/control PMT is located at the same position in the test stand for all of the PMT tests. 
So as not to disturb the experimental environment, LN2 is not added to the dewar during the data taking period. 
The tubes are illuminated by the fiber-connected diffuser above the liquid surface.
The pulser circuit and the LED are located sufficiently far away from the cryogenics to avoid 
temperature-dependent performance changes
The LED pulser frequency is set to 10~Hz for all measurements 
except the rate dependence test (Sec.~\ref{subsubsec:rate}). 
LED intensity is controlled by the circuit bias voltage; the intensity 
is not absolutely calibrated and the absolute intensity is unknown. 
However, the number of photoelectrons (PEs) from 
the photocathode can be determined from the data,   
either from the Gaussian approximated Poisson distribution or 
the single photoelectron response (SER) spectrum depending on the light intensity. 
Using this information, the PMT gains are measured.
The Tektronix TDS-5054B Digital Oscilloscope~\cite{oscilloscope} was used for data acquisition, and
a Dzero high voltage power supply with the CSS monitoring system~\cite{CSS} was used to provide and 
monitor the PMT high voltage.

\section{Results}

\subsection{Special tests}

\begin{figure}
\centering
\includegraphics[width=0.60\textwidth]{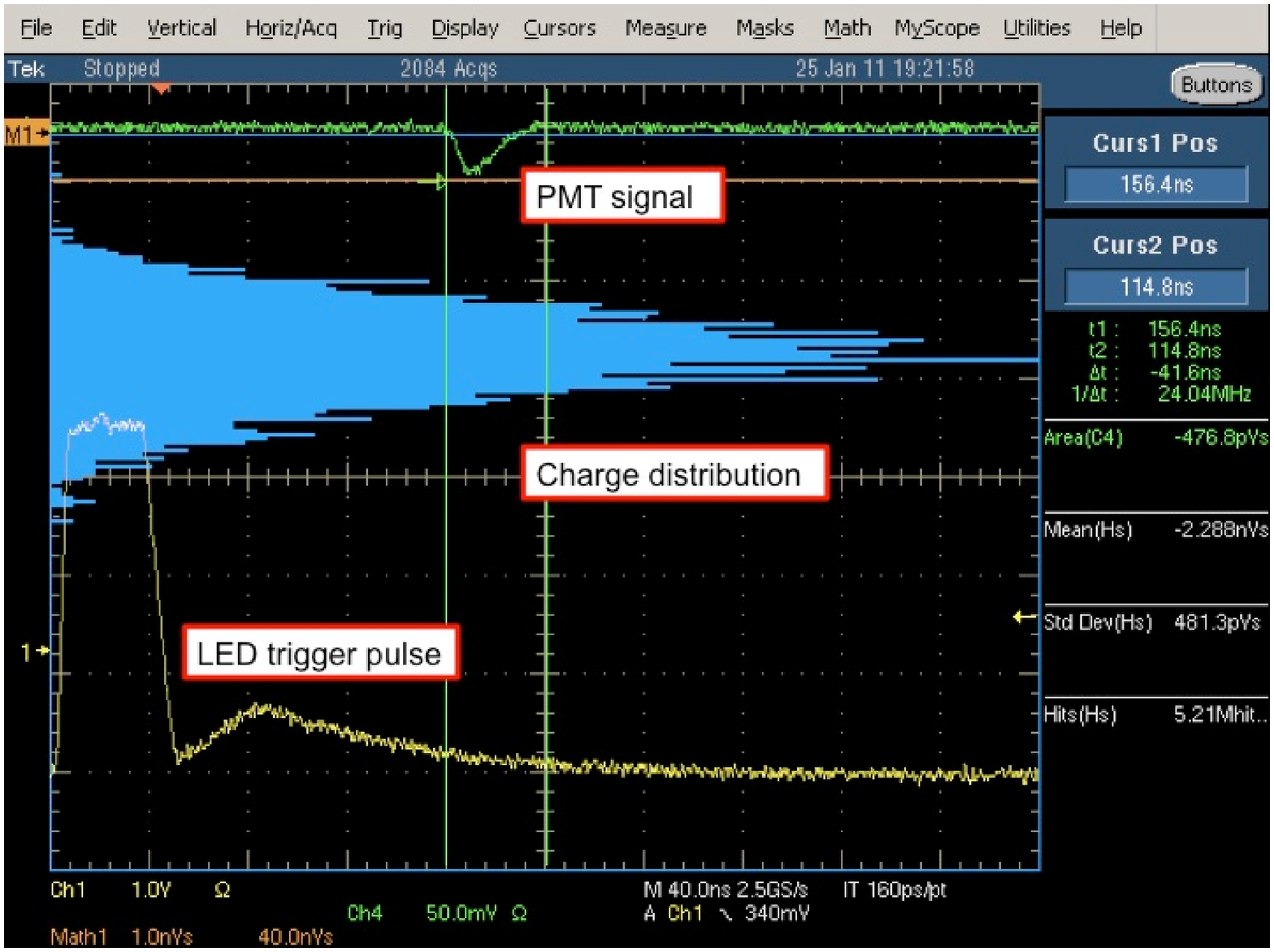}
\caption{\label{fig:poisson}
A typical screen shot from the TDS-5054B oscilloscope. 
The upper green trace shows the split signal pulse from the PMT, 
and the lower yellow trace is the trigger pulse for the LED. 
Only the rising edge of the trigger pulse is important to trigger the LED; 
the LED pulse is $\sim$10~ns, shorter than the trigger pulse. 
The blue histogram rotated $90^\circ$ displays the distribution of the integrated PMT signal which measures
the charge output from the PMT; this makes a symmetric Poisson distribution when 
light intensity is high enough ($>$5 PEs). 
}
\end{figure}

Selected PMTs are tested with a relatively high number (5 - 10) of PEs  to  
check a number of tube properties. 
If the LED intensity is high, 
we first find the number of PEs through the Poisson distribution of the charge;
the gain of the PMT is obtained from the averaged charge. 
Figure~\ref{fig:poisson} shows the typical plot of this method. 
The mean and the variance are read from the symmetric Poisson distribution of the charge 
(blue histogram in Fig.~\ref{fig:poisson}), 
and the square of the ratio allows the number of PEs to be extracted.
From the variation in the number of measured PEs, we estimate a 10\% error on this PE measurement. 
This gives 12\% error to the gain determination by this method. 
This approach is affected by any environmental effects 
(reflection of light inside the Dewar, for example) 
that broaden the charge distribution. 
As a result, even though this method does not involve any fit calculations and therefore yields a quick result,
it is not considered to be as accurate as the gain calculated from fitting the single electron 
response described below (Sec.~\ref{subsec:gain}). 

\subsubsection{Stability test\label{subsubsec:stability}}

\begin{figure}
\centering
\includegraphics[width=0.80\textwidth]{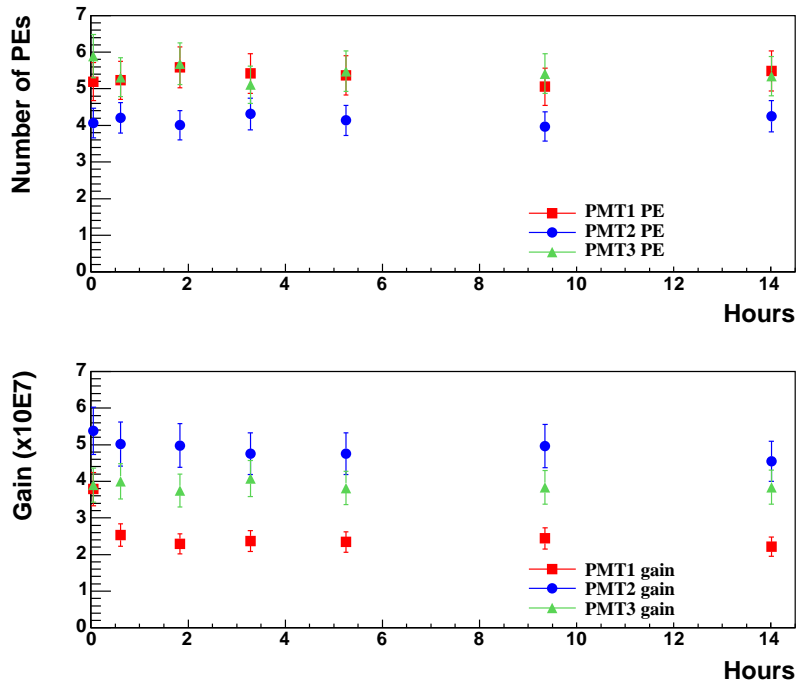}
\caption{\label{fig:TimeDep1}
PMT gain measurement stability test. 
High voltage is suddenly applied to cooled PMTs. 
The upper plots show the measured number of PEs; this demonstrates the long time stability of the environment.  
The lower plot shows measured gain. 
There is some anomalous behavior in the first $\sim$30~min, 
but the gain is stable thereafter.
}
\end{figure}

This test examines the stability of 
the photomultiplier gain by quickly raising the tube voltage 
from zero to operating voltage while the PMT is at cryogenic temperature. 
First, three PMTs are immersed in LN2 for more than three days, 
during which time the tubes are kept in a light-tight environment with no applied HV. 
The pulsed LED  illumined the tubes at low intensity, 
with each PMT receiving sufficient light to generate $\sim$5 PEs. 
The tube voltage was then quickly ramped to an operating level where all the tubes had roughly the same gain ($\sim10^{7}$). 
Figure~\ref{fig:TimeDep1} shows the result of three PMTs measured simultaneously for this test.  
The upper plot shows the variation in the number of measured PEs as a function of a time. 
As noted above, the number of PEs is assigned a 10\% error, which covers the observed variation. 
The number of measured PEs varies slightly with the location of the PMT in the fixture, but 
the number is stable for each tube over the time period shown. 
This stability demonstrates that the test environment does not change appreciably on this time scale, 
including both the liquid nitrogen level (changes in which would alter light propagation inside the Dewar) 
and the LED pulser system (especially the LED intensity).

The lower plot in Figure~\ref{fig:TimeDep1} shows the variation in the PMT gains. 
The gains change quite a bit in the first $\sim$30~min, 
but after that, they are stable.
This observation confirms that changing the PMT voltage while the tubes are at cryogenic temperatures 
does not have any particular effect on tube performance.

\subsubsection{Gain change with cooling time\label{subsubsec:cooling}}

\begin{figure}
\centering
\includegraphics[width=0.80\textwidth]{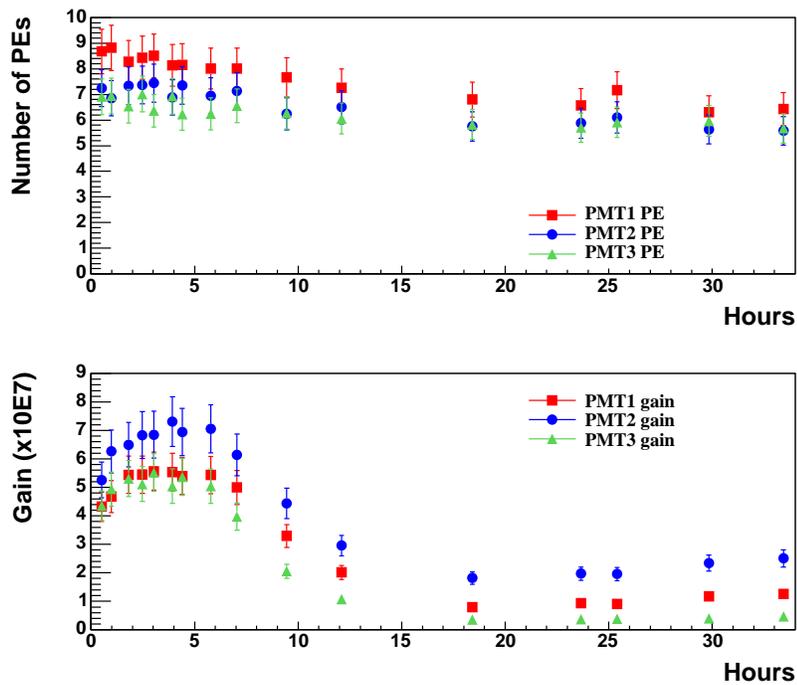}
\caption{\label{fig:TimeDep2}
 PMT cooling  test. 
The drift in the number of measured PEs suggests that the photocathodes 
are excited before the measurement by exposure to ambient light.
The bottom plot shows the gain variation with time. 
The gain drops sharply after $\sim$5 hours, 
and reaches a stable value later.
}
\end{figure}

The next gain stability test examines the variations when the PMT is not initially cooled for several days.
Figure~\ref{fig:TimeDep2} shows the results of this study; in this case,
the tube is brought up to operating voltage immediately after LN2 immersion.  
As with the previous test, data for three PMTs are collected simultaneously to check consistency.
Data taking starts after the initial post-immersion boiling of the LN2 subsides ($\sim$15~min). 
Note the structure of the test stand does not permit keeping PMTs completely in a dark environment 
before immersion; as a result, the data show the combination of two effects: the actual PMT cooling, 
and the excitation of the photocathode due to pre-immersion light exposure. 
This makes it difficult to interpret the drift of measured PEs. 
Despite this effect, however, 
the dramatic change in PMT gain can clearly be observed.
The PMT gain remains high in the first $\sim$5 hours after immersion, 
and then suddenly drops by more than a factor of two, coming to a stable value with a small drift.
This effect is attributed to the tube's interior coming to 
thermal equilibrium; to insure that the tube is completely
cooled, all other tests described in this note will immerse the tubes for three days before proceeding.

\subsubsection{Linearity test}

\begin{figure}
\centering
\includegraphics[width=0.60\textwidth]{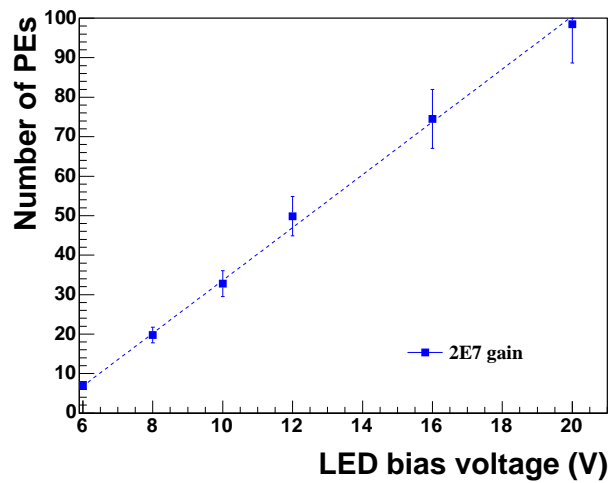}
\caption{\label{fig:linearity}
A typical linearity test plot. The number of PEs is measured as a function of LED bias voltage, 
which affects the light output. 
Non-linearity would be evident in narrower Poisson distributions at higher light levels, 
which would make the number of visible PEs higher.
}
\end{figure}

This section details how the PMT response varies as a function of the LED intensity; 
this effect is tested using several PMTs. 
The LED output is first monitored by a PMT set to low gain, in order to find the dynamic range of the LED. 
When the LED bias voltage is 0 to 20 V, the LED intensity shows a linear dependence on the bias voltage. 
This LED is then used to measure the number of PEs observed in the PMT as a function of the LED bias voltage. 
Figure~\ref{fig:linearity} shows a typical result. 
The maximum attainable number of PEs for the cryogenic temperature test is around 50 to 100, 
Within this range and the expected MicroBooNE PMT gain 
($\sim3\times10^7$, discussed in Sec.~\ref{subsec:gain}), 
all tested PMTs show good linearity; this provides a measure of confidence that the MicroBooNE PMTs
will perform in a linear fashion up to $\sim$100~PEs 
in the cryogenic environment. 

\subsubsection{Rate dependence of PMT gain\label{subsubsec:rate}}

\begin{figure}
\centering
\includegraphics[width=0.80\textwidth]{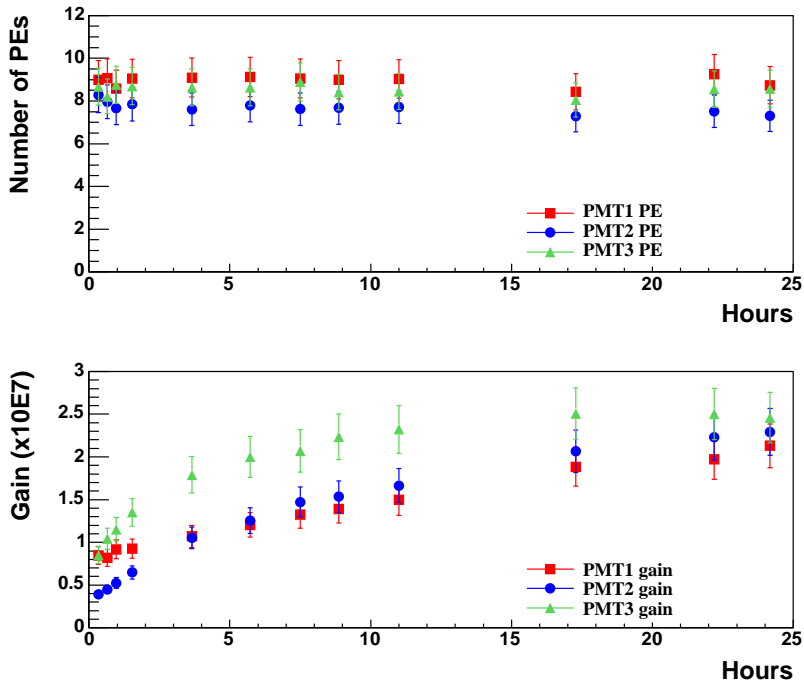}
\caption{\label{fig:RateDep1}
Repeating the stability test with higher LED pulser frequency (10~kHz, instead of 10~Hz). 
Note that PMT gains change on a long time scale,
and it takes $\sim$one day or more to reach the gain plateau. 
}
\end{figure}

\begin{figure}
\centering
\includegraphics[width=0.80\textwidth]{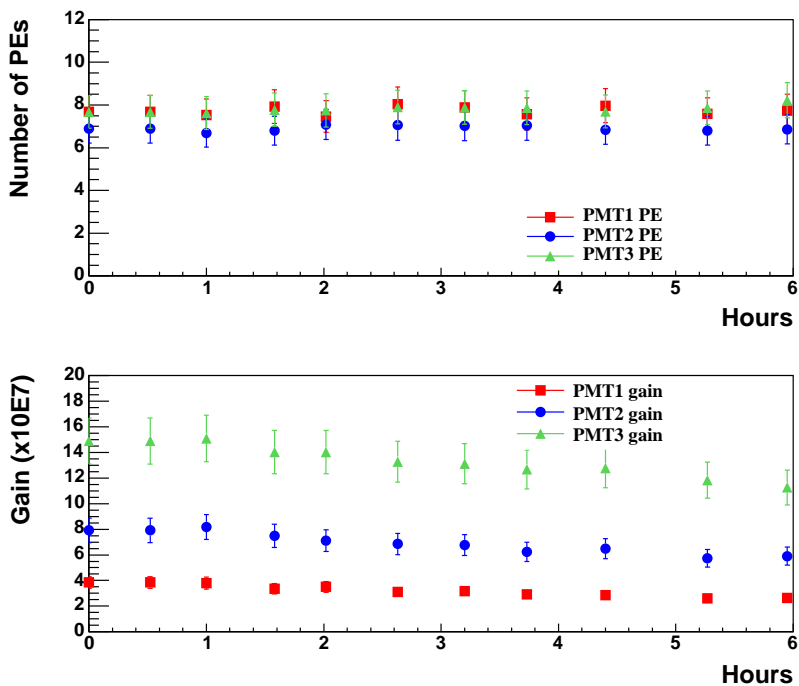}
\caption{\label{fig:RateDep2}
PMT gain drift by changing LED frequency. 
Here, LED frequency is changed from 10~Hz to 10~kHz over 6 hours. 
Note that during data taking period measured PEs are constant for all PMTs.
}
\end{figure}

The PMT rate test shows that a high rate of LED pulses affect tube performance. 
Figure~\ref{fig:RateDep1} is the same test as performed in Sec.~\ref{subsubsec:stability} 
(Fig.~\ref{fig:TimeDep1}), 
but in this instance the pulser driving the LED is running at higher frequency (10~kHz).  
The gains of the PMTs change on a long time scale after being brought to operating voltage. 
The number of measured PEs is observed to be stable, and as with the previous  stability test 
there appears to be no change in the test environment. 
The gains increase very slowly, and it takes over a day to reach the plateau. 
Also this gain plateau is lower than that measured at low frequency.

A second related test examined the effect of changing the LED pulser frequency. 
Figure~\ref{fig:RateDep2} is an example of such effect. 
Here, the LED frequency is changed from 10~Hz to 10~kHz over 6 hours. 
The LED frequency is changed 5 times, 10 to 100, 100 to 1,000, 1,000 to 2,000, 
2,000 to 4,000 then 4,000 to 10,000 in every hour. 
In this range, the PMT gain is 10-20\% suppressed. 
Also, when LED frequency is high, 
PMT gain shows both a long time-scale drift and a hysteresis  
over the changing of the LED intensity and/or the PMT HV.  
All rate dependencies of PMT gains are present only for the LN2 tests, 
and common to all tested PMTs; specific details, however,  may depend 
on the individual characteristics of each PMT.  
We also note that we do not observe any of 
these phenomena with room temperature running with 10~kHz LED. 
This is a confirmation that the 10~kHz 
is not considered the fast signal for our PMT and base. 
One can speculate that dynodes later in the chain may lack sufficient electrons to respond when 
the pulse frequency is high in the cold environment 
due to the low mobility of electrons, 
but the real cause is presently unknown. 
This has some implications for PMT operation in a high rate background, 
such as a LAr detector running on the surface and exposed to cosmic rays.
Under such conditions, the PMTs would not achieve their optimal gain, 
even though the gain would reach a plateau at some point.
This test suggests that this optimal plateau may be somewhat unstable, 
therefore running cryogenic PMTs in high rate environments may require special attention.

\subsection{MicroBooNE PMT gain measurement\label{subsec:gain}}

\begin{figure}
\centering
\includegraphics[width=0.60\textwidth]{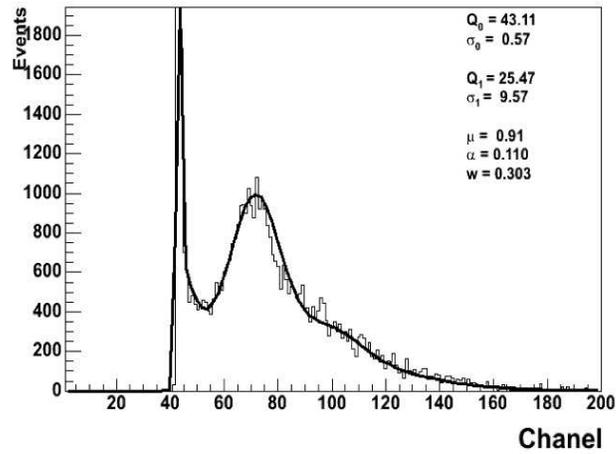}
\caption{\label{fig:SER}
A typical screen shot for the SER fit, performed for all tested PMTs 
from 900 to 1800 V. 
The pedestal, single photoelectron peak, and a small two PE peak can be seen. 
The fit simultaneously finds seven parameters  of the single photoelectron response; for details, see
Ref.~\cite{YiChen}. 
}
\end{figure}

The measurement of PMT gain runs the LED at low intensity (producing 1-2~PEs), allowing
the PMT gain to be found from the separation of the single PE peak from the pedestal. 
This is accomplished with greater accuracy by fitting the single photoelectron response plot, 
using a procedure described in Ref.~\cite{YiChen}.
This approach deals correctly with the two types of noise
(signal broadening and spurious small signal) and 
allows absolute calibration. 
The SER is recorded in 100~V steps from 900 to 1800 V and fitting this response allows 
the gains of all 30 PMTs to be measured, both in air and in LN2. 
Figure~\ref{fig:SER} shows the typical fit. 
The procedure simultaneously fits seven SER parameters, 
including the location of the pedestal and the single PE peak;
the separation of these two parameters measures the PMT gain. 

\begin{figure}
\centering
\includegraphics[width=0.80\textwidth]{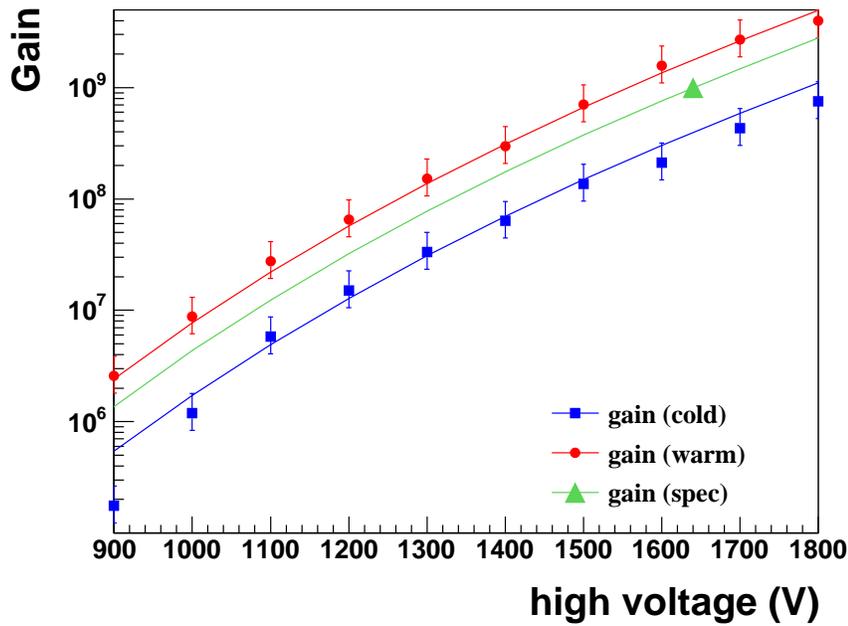}
\caption{\label{fig:typicalgain}
A typical gain plot for a MicroBooNE PMT. 
All of them are overlaid with 11th power law.
The blue square markers are for the LN2 running, 
the circle markers are for room temperature,
and the green triangle marker is from the manufacturer's specification sheet. 
}
\end{figure}

The MicroBooNE electronics design requires $\sim3\times10^7$ gain, 
so the one of the goals of the PMT test is to find the  HV needed to achieve this gain for each PMT.
Figure~\ref{fig:typicalgain} shows a typical gain of a PMT. 
The blue square markers are data taken at cryogenic (LN2) temperature, 
and the red circle markers are from the room temperature run.
The green triangle markers are for the gain from the manufacturer's specifications sheet. 
The specifications only provide one HV value to achieve the $10^9$ gain, 
so the rest of the points  are extrapolated
by assuming that the gain goes as the voltage to the eleventh power;
so the rest of the points  are extrapolated
the exponent is chosen from the data (red and blue curves in Fig.~\ref{fig:typicalgain}). 
The error is estimated from the multiple measurements made on the calibration PMT. 
The calibration PMT is fixed at the same location during all tests, 
and it is immersed in LN2 and warmed up multiple times to monitor  
changes in the test environment. 

\begin{figure}
\centering
\includegraphics[width=0.60\textwidth]{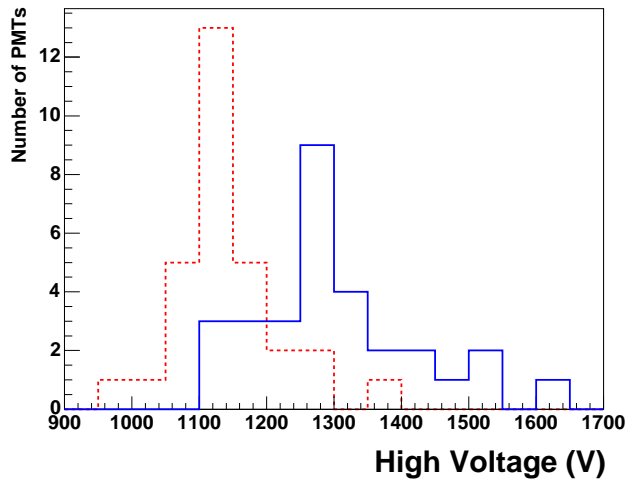}
\caption{\label{fig:optimalHV}
Voltage distributions for the 30 PMTs set to provide $3\times10^7$ gain. 
The blue solid histogram id the distribution of the voltages at LN2 temperature,
and the red dashed histogram is the distribution of the voltages at room temperature. 
}
\end{figure}

The slope of the measured gains can be seen to follow 
a simple power law, with some variations at the
the low and high voltage ends of the distribution.
The gain does however follow this power law in the region of interest, 
where a PMT achieves a $\sim3\times10^7$ gain. 
This allows an operating HV to be extrapolated from neighboring values by 
fitting to a power law.
Most of the PMTs show gains 10 to 30$\%$ higher than the $10^9$ 
at the HV values given in the manufacturer's specifications.
The reason for this is presently unknown. 
Also, as expected, the gain in LN2 is lower than the gain at room temperature.
Typically, the cold gain is $\sim$10-50\% of the warm gain. 
This result  is consistent with other measurements~\cite{ICARUS,Bueno}. 
Figure~\ref{fig:optimalHV} shows the summary of optimal HV values to 
achieve $3\times10^7$ gain from all PMTs. 
It is easy to compensate for the diminution in the PMT gain caused by running at cryogenic temperatures
by increasing the tube voltage  $\sim$200~V. 

\subsection{MicroBooNE PMT dark current measurement}

\begin{figure}
\centering
\includegraphics[width=0.80\textwidth]{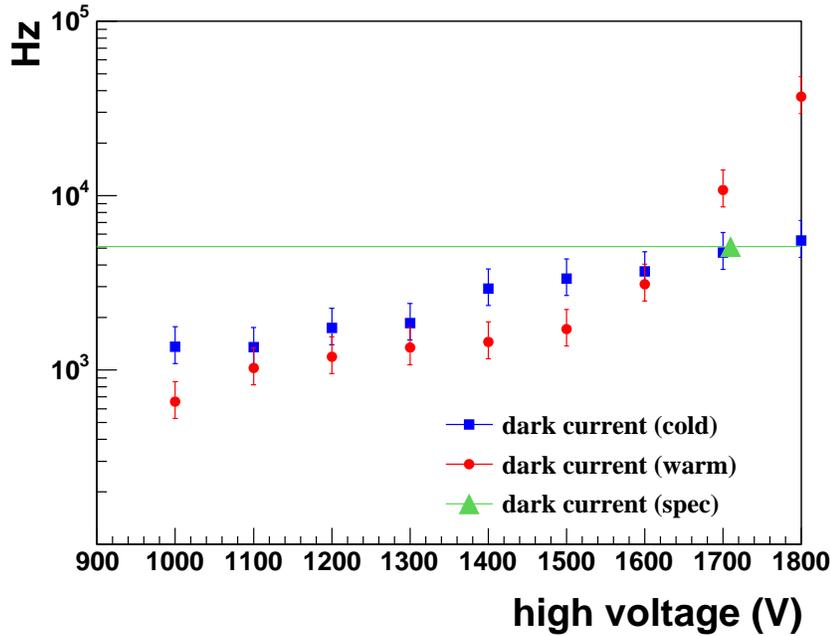}
\caption{\label{fig:typicalnoise}
A typical dark current plot for a MicroBooNE PMT. 
The blue square markers are for the LN2 running,
and the red circle markers are for room temperature.
The green triangle marker shows the dark current value 
at $10^9$ gain at optimal HV from the manufacturer's specifications. 
}
\end{figure}

The dark current is measured for all PMTs in air and for half of the PMTs in LN2. 
It is known that the dark current at cryogenic temperature is higher than the room temperature 
for platinum under layered cryogenic PMTs~\cite{Hans1,McKinsey, Hans2}. 
The dark current is defined as the rate of  spontaneous single PE signals without a light source, 
at a level above a  3~mV (5~mV) threshold for the cold (warm) runs. 
This is the typical single PE signal height for PMTs when the gains are low.  
This definition cannot be rigorously correct, 
because each PMT has a different single PE pulse height;
some PMTs may only barely achieve this threshold when the HV is low, 
or for high gain PMTs, there may be too many $<1$ PE signals which pass this threshold. 
Figure~\ref{fig:typicalnoise} shows a typical dark current plot from one PMT. 
Again, the blue square markers are for the cold run, 
and the red circle markers are for the warm run. 
The green triangle marker shows the dark current value from the manufacturer's specifications. 
As before, the specifications provide only a single point at the HV value where 
the gain of the PMT achieves $10^9$. Not knowing the trend of the dark current values made it impossible to
extrapolate additional points. 
The error is estimated from the variation in the calibration PMT data.  
The PMT shows a dark current count plateau over a wide range, 
where the gain increases exponentially, but the dark current increases only linearly. 
This effect holds true for all PMTs, leading to the above definition of the dark current. 
Most PMTs show a breakdown of the plateau around 1600-1700~V at room temperature, but
all dark current measurements in LN2 show plateau behavior up  to 1800~V; this is consistent with Ref.~\cite{Hans2}. 

\section{Conclusion}

This paper reports the testing procedures and results for the MicroBooNE photomultiplier tubes.
The PMT base has been designed and tested, bases have soldered to all of the PMTs, 
and all tubes have been confirmed as working in a cryogenic (LN2) environment.
The cryogenic PMT test stand was designed utilizing a large  Dewar and
was found to provide a test environment stable on a long time scale.
Several characteristic behaviors of a sample of the cryogenic PMTs were measured, as were
 gains and dark currents for  all PMTs.
The PMT gains were found  to be lower at cryogenic temperatures, 
but increasing the tube voltage by $\sim$200 V compensates for this decrease.

\acknowledgments

The authors thank the National Science Foundation (NSF-PHY-0847843 and 
PHY-1000214) for financial support. 
We thank Yi Chen of the Brookhaven National Laboratory for sharing the fit code, 
and Stephen Pordes, Linda Bagby, Sten Hansen, Cary Kendziora, William Miner, Kelly Hardin, 
and Nathan Bremer of the Fermi National Accelerator Laboratory for helping to build the system. 
We also thank to Flavio Cavanna, Steven Linden, and Vasilli Papavassiliou for 
their careful readings of this manuscript.

\end{document}